\documentclass[useAMS]{mn2e}
\usepackage{graphicx}
\usepackage{epsfig}
\usepackage{amsmath} 
\usepackage{amssymb}

\def\ltsima{$\; \buildrel < \over \sim \;$}
\def\lsim{\lower.5ex\hbox{\ltsima}}
\def\gtsima{$\; \buildrel > \over \sim \;$}
\def\gsim{\lower.5ex\hbox{\gtsima}}
\def\mdot {\dot M}
\newcommand{\be}{\begin{equation}}
\newcommand{\en}{\end{equation}}
\newcommand{\ergs}{\rm \ erg \; s^{-1}}

\def\msole {~M_{\odot}}

\begin{document}
\title[Engulfing of PSR J1023+0038]{Engulfing a radio pulsar: the case of PSR J1023+0038}

\author[F. Coti Zelati et al.]{F. Coti Zelati,$^{1,2}$\thanks{E-mail: francesco.cotizelati@brera.inaf.it} 
M. C. Baglio,$^{1,2}$ S. Campana,$^2$ P. D'Avanzo,$^2$ P. Goldoni,$^3$ \newauthor N. Masetti,$^4$
T. Mu\~{n}oz-Darias,$^5$ S. Covino,$^2$ R. P. Fender,$^5$ E. Jim\'{e}nez Bail\'{o}n,$^6$ \newauthor
H. Ot\'{i}-Floranes,$^6$ E. Palazzi$^4$ and F. G. Ram\'{o}n-Fox$^6$\\
$^1$ Universit\`a dell'Insubria, Via Valleggio 11, I-22100 Como, Italy\\
$^2$ INAF - Osservatorio astronomico di Brera, Via Bianchi 46, I--23807, Merate (LC), Italy\\
$^3$ APC, Astroparticules et Cosmologie, Universit\'{e} Paris Diderot, CNRS/IN2P3, CEA/Irfu, Observatoire de Paris, Sorbonne Paris Cit\'{e}, 10,\\ rue Alice Domon et L\'{e}onie Duquet, 75205, Paris Cedex 13, France\\
$^4$ INAF - Istituto di Astrofisica Spaziale e Fisica Cosmica di Bologna, via Gobetti 101, I--40129 Bologna, Italy\\
$^5$ University of Oxford, Department of Physics, Astrophysics, Keble Road, Oxford, OX1 3RH, UK\\
$^6$ Instituto de Astronom\'{i}a, Universidad Nacional Aut\'{o}noma de M\'{e}xico, Apartado Postal 106, Ensenada B. C., 22800, Mexico\\
}
\date{Accepted 2014 July 30. Received 2014 July 8; in original form 2014 April 4}

\pagerange{\pageref{firstpage}--\pageref{lastpage}} \pubyear{2014}

\maketitle

\label{firstpage}

\begin{abstract}
The binary millisecond radio pulsar PSR J1023+0038 has been recently the subject of multi-wavelength monitoring campaigns 
which revealed that an accretion disc has formed around the neutron star (since 2013 June). We present here the results of X-ray 
and UV observations carried out by the {\it Swift} satellite between 2013 October and 2014 May, and of optical and NIR 
observations performed with the REM telescope, the Liverpool Telescope, the 2.1-m telescope at the San Pedro 
M\'{a}rtir Observatory and the 1.52-m telescope at the Loiano observing station.  The X-ray spectrum is well described by an absorbed 
power law, which is softer than the previous quiescent epoch (up to 2013 June). The strong correlation between the X-ray 
and the UV emissions indicates that the same mechanism should be responsible for part of the
emission in these bands. Optical and infrared photometric observations show that the companion star is strongly irradiated. 
Double-peaked emission lines in the optical spectra provide compelling evidence for the presence of an outer accretion disc too. 
The spectral energy distribution from IR to X-rays is well modelled if the contributions from the companion, the disc and the intra-binary 
shock emission are all considered. Our extensive data set can be interpreted in terms of an engulfed radio pulsar: the radio pulsar is 
still active, but undetectable in the radio band due to a large amount of ionized material surrounding the compact object. X-rays and 
gamma-rays are produced in an intra-binary shock front between the relativistic pulsar wind and matter from the companion and an 
outer accretion disc. The intense spin-down power irradiates the disc and the companion star, accounting for the UV and optical emissions.
\end{abstract}


\section{Introduction}
FIRST J102347.6+003841 (hereafter J1023) was discovered by Bond et al. (2002) in the radio band and initially classified as 
a magnetic cataclysmic variable. Optical studies revealed signs for the presence of an accretion disc in 2001 (Szkody et al. 
2003; Wang et al. 2009a), which led to identify this system as a neutron star low-mass X-ray binary (LMXB; 
Thorstensen \& Armstrong 2005; Homer et al. 2006). No enhanced X-ray emission was reported. The discovery of a 1.69-ms 
radio pulsar, PSR J1023+0038, in a 4.75-h binary system (Archibald et al. 2009) made J1023 the first system showing the 
potential to alternate its radio pulsar phase, powered by rotation, to an X-ray state, powered by accretion and typical of LMXBs. 
The recycling scenario of neutron stars in LMXBs in fact does not necessarily proceed as a stable phase: a large majority of 
LMXBs are indeed transients and when the neutron star is already spinning at the ms level, accretion episodes (outbursts) can 
alternate to radio ejection phases in the same system (Campana et al. 1998; Burderi et al. 2001). This picture has been successfully 
confirmed by IGR J18245--2452 in the globular cluster M28, which presented itself as a millisecond radio pulsar, an accreting 
millisecond X-ray pulsar and again as a millisecond radio pulsar on a time-scale of a few weeks (Papitto et al. 2013). A similar system 
is the LMXB XSS J12270--4859: radio, optical and X-ray observations suggested that this system has recently switched to a millisecond 
radio pulsar phase (Bassa et al. 2014; Bogdanov et al. 2014). The detection of 1.69-ms radio pulsations provided compelling evidence 
of this state change (Roy et al. 2014).

Since its discovery in 2007 June, J1023 has been regularly monitored with several radio facilities. A long-term 
radio monitoring campaign from mid-2008 to mid-2012 revealed eclipses, interpreted as the result of magnetic activity of the 
companion (Archibald et al. 2013). J1023 was detected with the Westerbork Synthesis Radio Telescope on 2013 June 15, 
but it was not detected by the Lovell Telescope at Jodrell Bank eight days later (Stappers et al. 2014). Deeper 
observations at the same or higher frequencies confirmed this disappearance (Stappers et al. 2014). During this period (2013 
July--August), J1023 remained below the sensitivity threshold of the Burst Alert Telescope hard X-ray monitoring instrument on
board {\it Swift}. This results in a 15--50 keV upper limit for the luminosity (for a source distance of 1368 pc\footnote{The 
source distance was determined with a 3 per cent error by measuring the parallax of the system.}; Deller et al. 2012) of 
$\sim 10^{35}\ergs$ (Stappers et al. 2014) and rules out a major accretion outburst from this system. J1023 remained undetected 
in the radio also during an observation in 2013 October (Patruno et al. 2014).

As many other millisecond radio pulsars, J1023 is a known gamma-ray emitter (Tam et al. 2010). The {\it Fermi} satellite 
revealed an increase in the GeV flux (100 MeV--300 GeV) by a factor of $\sim 5 - 7$ compared to the long-term average during 
the periods 2013 Jun 21--Aug 17 and Aug 17--Sep 15, respectively (Stappers et al. 2014). Despite some variability at 
time-scales of a few days to a few weeks, the gamma-ray flux remained at a high level for most of the time until, at least,
2013 November (Takata et al. 2014).

In the X-ray band, J1023 has been observed several times in 2013 by {\it Swift}/XRT. The system appears to be much brighter
on October 18 than on June 10 and 12, reaching a 0.3--10 keV luminosity of $2.5\times 10^{33}\ergs$ 
(Kong 2013; Patruno et al. 2014), a factor $\sim 30$ brighter than its quiescent level ($\sim 9\times 10^{31}\ergs$; 
Archibald et al. 2010; Bogdanov et al. 2011). Moreover, evidence for aperiodic variability is found on short time-scales, 
down to $\sim10$ s. {\it NuSTAR} observations also revealed an increase in the 3--79 keV luminosity of the system, from
$\sim7.4\times 10^{32}\ergs$ between June 10--12 to $\sim5.8\times 10^{33}\ergs$ and up to $1.2\times 10^{34}\ergs$ 
between October 19--21 (Tendulkar et al. 2014).

In the optical, the system was $\sim 1$ mag brighter on October 16-23 than in June and somewhat bluer (Halpern et al. 2013). 
This is confirmed by UV observations gathered with {\it Swift}/UVOT, since an increase by $\sim 3.5$ mag is observed in the 
UVW1 filter (Patruno et al. 2014). Optical spectroscopy performed between 2013 October 22 and December 16 shows the presence 
of several broad emission lines such as H$\alpha$, H$_{eI}$ $\lambda$5876, and H$_{eI}$ $\lambda$6678 (Halpern et al. 2013; Linares et al. 
2014; Takata et al. 2014). The H$\alpha$ line is strong with an equivalent width of $\sim 24$\AA\, and double-peaked, 
indicating the presence of an accretion disc. These values are similar to those observed during the 2001 optically active 
phase (Wang et al. 2009a).

In this paper, we confirm and improve this picture with new observational data. We report on a much larger number of {\it Swift} observations in 
the X-ray and UV/optical bands in Section 2. We focus on new optical and infrared photometry of this system in Section 3. 
We report the results of new optical spectroscopy in Section 4. We derive a strictly contemporaneous spectral energy distribution (SED) in Section 5.
We put forward a theoretical interpretation based on a completely enshrouded radio pulsar in Section 6. Conclusions are drawn in Section 7.

\section{Swift observations}
The {\it Swift} satellite (Gehrels et al. 2004) observed J1023 51 times between 2013 October 18 and 2014 May 12.
The observations were carried out both in the soft X-ray (0.3--10 keV) and in the UV/optical bands with the XRT and the
UVOT on board {\it Swift}, respectively. We report the main characteristics of the observations in Table \ref{log}.
The first 10 {\it Swift} observations were already analysed by Takata et al. (2014).

\begin{table*}
\caption{{\it Swift} XRT and UVOT observational log. The observation marked with $^*$ was already analysed by Patruno et al. (2014) and Takata et al. (2014). 
Observations marked with $^{**}$ were already analysed by Takata et al. (2014).}
\label{log}
\begin{center}
\begin{tabular}{lcccc}
\hline
Obs. ID     			& Date        		& XRT exp. 	& UVOT filter (mode)			& XRT count rate\\
            			&             	  		& (ks)       		&      		      				& (counts s$^{-1}$)\\
\hline
00080035003$^*$	& 2013 Oct 18   	& 10.0		& U, UVW1 (img)			& $0.28\pm0.01$\\
00033012001$^{**}$ & 2013 Oct 31 	& 1.9			& U (img)					& $0.20\pm0.01$\\
00033012002$^{**}$ & 2013 Nov 04 	& 2.1			& U (img)					& $0.28\pm0.01$\\
00033012003$^{**}$ & 2013 Nov 06 	& 1.0			& UVW1 (evt)				& $0.21\pm0.02$\\
00033012004$^{**}$ & 2013 Nov 07 	& 1.0			& UVW1 (evt)				& $0.19\pm0.02$\\
00033012005$^{**}$ & 2013 Nov 08 	& 1.3			& UVW1 (evt)				& $0.36\pm0.02$\\
00033012007$^{**}$ & 2013 Nov 10 	& 1.1			& UVW1 (evt)				& $0.21\pm0.02$\\
00033012008$^{**}$ & 2013 Nov 11 	& 1.2			& UVW1 (evt)				& $0.16\pm0.01$\\
00033012009$^{**}$ & 2013 Nov 12 	& 2.3			& UVW1 (evt)				& $0.23\pm0.01$\\
00033012010$^{**}$ & 2013 Nov 13 	& 1.1			& UVW1 (evt)				& $0.14\pm0.01$\\
00033012011 		  & 2013 Nov 14 	& 1.1			& UVW1 (evt)				& $0.23\pm0.02$\\
00033012012 		& 2013 Nov 15 		& 1.1			& UVW1 (evt)				& $0.19\pm0.01$\\
00033012013 		& 2013 Nov 16 		& 2.0			& UVW1 (evt)				& $0.23\pm0.01$\\
00033012014 		& 2013 Nov 17 		& 1.0			& UVW1 (evt)				& $0.27\pm0.02$\\
00033012015 		& 2013 Nov 18 		& 1.0			& UVW1 (evt)				& $0.29\pm0.02$\\
00033012016 		& 2013 Nov 19 		& 1.0			& UVW1 (evt)				& $0.30\pm0.02$\\
00033012017 		& 2013 Nov 20 		& 1.8			& U (img)					& $0.37\pm0.01$\\
00033012018 		& 2013 Nov 23 		& 1.9			& UVW1, UVW2, UVM2 (img)	& $0.29\pm0.02$\\
00033012019 		& 2013 Nov 28 		& 2.1			& U (img)					& $0.30\pm0.01$\\
00033012020 		& 2013 Nov 30 		& 4.5			& UVM2 (evt)				& $0.29\pm0.01$\\
00033012021 		& 2013 Dec 02 		& 0.7			& UVW1, UVW2, UVM2 (img)	& $0.05\pm0.01$\\
00033012022 		& 2013 Dec 11 		& 3.0			& UVW1 (img)				& $0.30\pm0.02$\\
00033012023 		& 2013 Dec 14 		& 1.0			& UVW1, UVW2, UVM2 (img)	& $0.24\pm0.01$\\
00033012024 		& 2013 Dec 12 		& 2.7			& UVW1 (img)				& $0.20\pm0.01$\\
00033012025 		& 2013 Dec 17 		& 1.8			& UVW1 (img)				& $0.17\pm0.01$\\
00033012026 		& 2013 Dec 21 		& 2.6			& UVW1 (img)				& $0.20\pm0.03$\\
00033012027 		& 2013 Dec 25 		& 2.8			& UVW1, UVW2, UVM2 (img)	& $0.26\pm0.01$\\
00033012028 		& 2014 Jan 02 		& 0.5			& UVW1, UVW2, UVM2 (img)	& $0.28\pm0.01$\\
00033012029 		& 2014 Jan 12 		& 2.8			& UVW1, UVW2, UVM2 (img)	& $0.22\pm0.01$\\
00033012030 		& 2014 Jan 16 		& 2.0			& UVW2 (img)				& $0.24\pm0.01$\\
00033012032		& 2014 Jan 22		& 0.6			& UVW2, UVM2 (img)		& $0.16\pm0.02$\\
00033012033		& 2014 Jan 24		& 3.0			& UVW2 (img)				& $0.35\pm0.01$\\
00033012034		& 2014 Jan 27		& 3.1			& U (img) 					& $0.25\pm0.01$\\
00033012036		& 2014 Feb 09		& 3.3			& UVW2 (img)				& $0.20\pm0.01$\\
00033012037		& 2014 Feb 11		& 1.0			& UVW1, UVW2, UVM2 (img)	& $0.24\pm0.02$\\ 
00033012038		& 2014 Feb 13		& 2.3			& UVW2 (img)				& $0.22\pm0.01$\\
00033012039		& 2014 Feb 19		& 2.8			& UVW1 (img) 				& $0.26\pm0.01$\\
00033012040		& 2014 Feb 21		& 0.4			& UVW1, UVW2, UVM2 (img)	& $0.21\pm0.03$\\ 
00033012041		& 2014 Feb 27		& 0.5			& UVW1 (img)				& $0.64\pm0.05$\\
00033012042		& 2014 Mar 03		& 1.0			& UVW1, UVW2, UVM2 (img)	& $0.61\pm0.04$\\
00033012043		& 2014 Mar 05		& 1.6			& UVW2 (img) 				& $0.27\pm0.01$\\
00033012044		& 2014 Mar 11		& 2.4			& UVW1 (img) 				& $0.54\pm0.03$\\
00033012046		& 2014 Mar 12		& 1.8			& U (img) 					& $0.21\pm0.01$\\
00033012047		& 2014 Mar 23		& 1.0			& UVW1, UVW2, UVM2 (img)	& $0.21\pm0.02$\\ 
00033012049		& 2014 Apr 13		& 0.08		& UVW2 (img)				& $0.20\pm0.07$\\
00033012050		& 2014 Apr 16		& 0.9			& UVW1, UVW2, UVM2 (img)	& $0.28\pm0.02$\\
00033012051		& 2014 Apr 22		& 1.0			& UVW1, UVW2, UVM2 (img)	& $0.18\pm0.01$\\
00033012052		& 2014 Apr 25		& 2.1			& UVW2 (img)				& $0.28\pm0.01$\\
00033012053		& 2014 Apr 27		& 3.0			& U, UVW1, UVW2 (img)		& $0.19\pm0.01$\\
00033012054		& 2014 May 02		& 0.8			& UVW1, UVW2, UVM2 (img)	& $0.33\pm0.02$\\
00033012055		& 2014 May 12		& 1.0			& UVW1, UVW2, UVM2 (img)	& $0.24\pm0.01$\\
\hline
\end{tabular}
\end{center}
\end{table*}

\subsection{X-ray data}
X-ray data were collected by the XRT in photon counting mode (2.5-s time resolution). We processed the data using the 
default parameter settings with \textsc{xrtpipeline} (v. 0.12.8) and determined the count rates through the `sosta' command of
\textsc{image} (v. 4.5.1). We downloaded the spectral files using the {\it Swift}/XRT data products generator (Evans et al. 2009). 
We then assigned the latest version of the calibration files available in 2014 May to the spectral files and we grouped the
source spectra to have at least 20 counts per bin. Spectra were then analysed with the \textsc{xspec} (v. 12.8.1) spectral fitting 
package (Arnaud 1996) using the $\chi^{2}$ statistics. In all the fits we took into account the effects of interstellar 
absorption through the \textsc{tbabs} model with cross-sections from Verner et al. (1996) and abundances 
from Wilms, Allen \& McCray (2000). We fit the overall spectrum (i.e. the sum of the spectra of all the 51 
pointings, corresponding to a total exposure of 92.8 ks) with three different one-component spectral models: a power law, 
a blackbody, and a neutron star atmosphere model. Parameter errors have been computed with $\Delta \chi^2 = 2.706$, corresponding 
to 90 per cent confidence level for one parameter of interest.  

Taking the power law model as a baseline ($\chi^{2}_\nu$ = 0.98 for 788 degrees of freedom, d.o.f.), we reveal the presence of 
a non-negligible absorption ($N_H=[5.2\pm0.07]\times 10^{20}$ cm$^{-2}$; a fit with an unabsorbed power law yields an unacceptable 
$\chi^{2}_\nu$ = 1.12 for 789 d.o.f.).\footnote{An absorbed blackbody or a neutron star atmosphere model result in 
$\chi^{2}_\nu$ = 3.0 (788 d.o.f.) and $\chi^{2}_\nu$ = 4.8 (789 d.o.f.), respectively.} The photon index is $\Gamma=1.56\pm0.03$ and 
the mean unabsorbed 0.3--10 keV flux is $(1.15\pm0.02)\times 10^{-11}$ erg cm$^{-2}$ s$^{-1}$, corresponding to a luminosity of 
$(2.58\pm0.05)\times 10^{33}\ergs$. We also searched for possible iron line features by superimposing to the power law component 
a Gaussian model centred at 6.4 and 6.9 keV. We found upper limits of 35 and 47 eV for the equivalent width, respectively. 
The addition of a soft component is not statistically required.

To investigate spectral variability as a function of the X-ray flux, we separated the observations in two count rate ranges (below and above 
0.25 counts s$^{-1}$; see Table \ref{log}) and we summed together the corresponding spectra. We then fitted the two spectra separately 
with an absorbed power law ($\chi^{2}_\nu$ = 1.04 for 679 d.o.f. for the low-counting rate spectrum and $\chi^{2}_\nu$ = 0.86 for 739 d.o.f. for
the high-counting rate spectrum). The values obtained for the column density ($N_{H,L} = 5.4^{+1.3}_{-1.2} \times 10^{20}$ cm$^{-2}$; 
$N_{H,H} = 5.1^{+1.0}_{-0.9} \times 10^{20}$ cm$^{-2}$) and the photon index ($\Gamma_L=1.57\pm0.04$; $\Gamma_H=1.57^{+0.04}_{-0.03}$) 
are consistent with being constant within the errors. Indeed, changes in the power law normalization alone are enough to account for the observed 
spectral changes. We conclude that the spectral shape remains almost the same independently of flux variations.

\subsection{X-ray flares}
\begin{figure}
\begin{center}
\includegraphics[width=5.8cm,angle=-90]{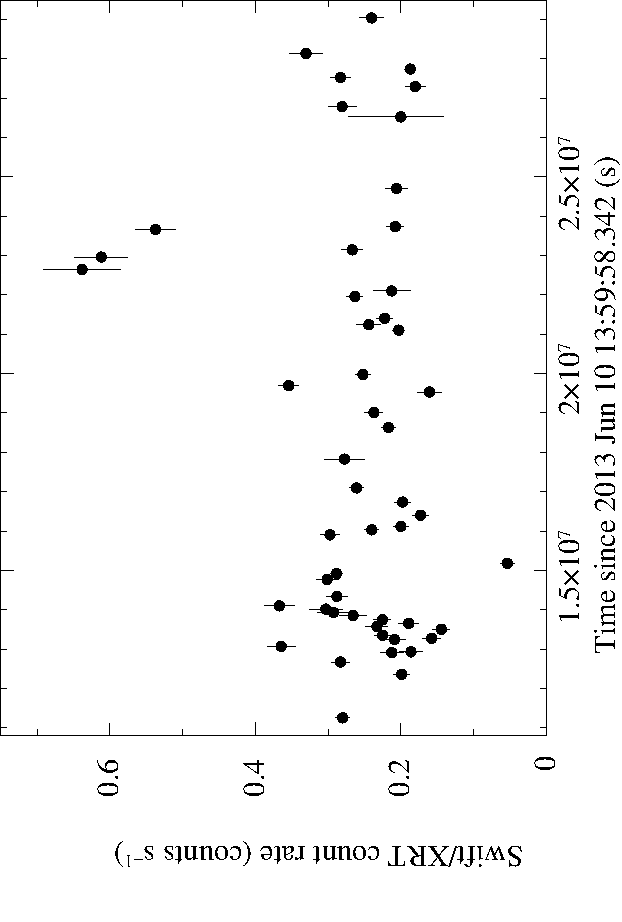}
\end{center}
\caption{0.3--10 keV light-curve of J1023 (2013 October 18--2014 May 12). Bin time equals one single observation.}
\label{LC}
\vskip -0.1truecm
\end{figure}

On 2014 February 27, the source X-ray count rate increased to $\sim0.64$ counts s$^{-1}$, i.e. a factor $\sim3$ higher than what was 
registered only 6 d before. The intensity decreased back on March 5, but the system was again at a high level on March 11. The following day the 
count rate returned back to roughly its average value (see Fig. \ref{LC} for the X-ray light-curve). These detections result in an upper limit of $\sim13$ 
d for the duration of such flares. X-ray spectral data of the three observations with the highest count rates (Obs. ID 00033012041, 00033012042, 
00033012044) are well described by an absorbed power law model ($\chi^{2}_\nu$ = 0.82 for 385 d.o.f., for a total exposure time of 3.8 ks) with hydrogen 
column density and photon index consistent with the average values ($N_H < 5.8\times 10^{20}$ cm$^{-2}$, $\Gamma=1.5\pm0.1$). The unabsorbed 
0.3--10 keV flux is ($2.5\pm0.2) \times 10^{-11}$ erg cm$^{-2}$ s$^{-1}$, which translates in to a luminosity $L_X = (5.6^{+0.5}_{-0.4})\times 10^{33}\ergs$, 
a factor $\sim2.2$ higher than the average value.

\begin{figure}
\begin{center}
\includegraphics[width=5.7cm,angle=-90]{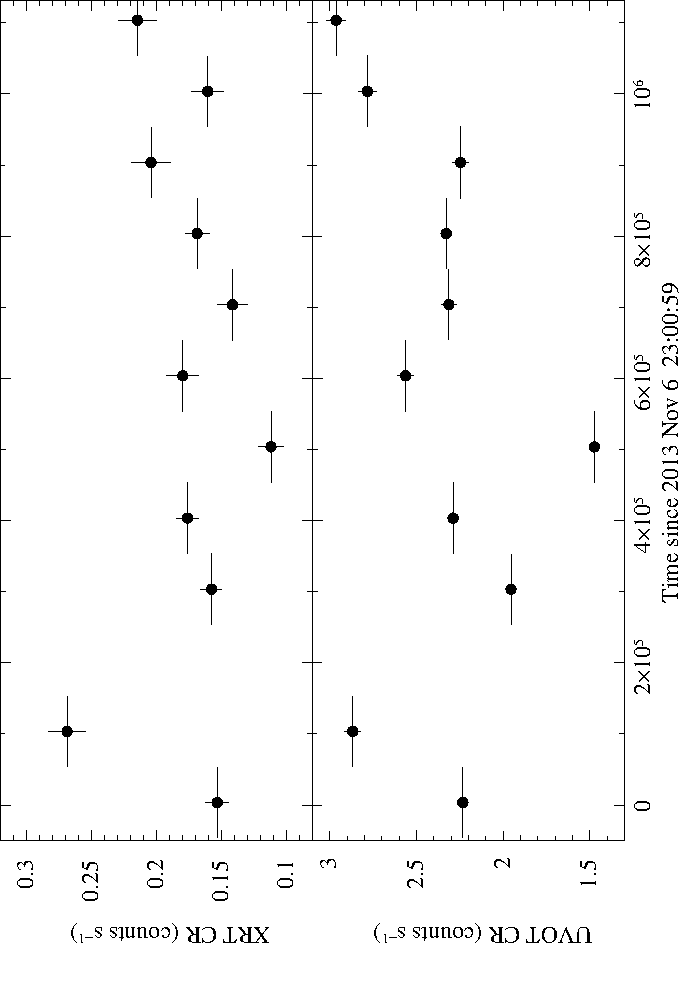}
\end{center}
\caption{0.3--10 keV and UVW1 light-curves of J1023 (2013 November 6--19). All data are grouped with a bin time of 10$^5$ s.}
\label{corr}
\vskip -0.1truecm
\end{figure}

\begin{table}
\caption{X-ray/UV variability and correlations in single {\it Swift} observations of J1023. A bin time of 50 s was adopted 
for each light-curve. Reported $\chi^{2}_\nu$ values are derived from fitting the light-curves with a constant. $P_s$ is the 
significance probability for the X-ray/UV correlation according to Spearman's correlation test.}
\label{varcorr}
\begin{center}
\begin{tabular}{c|cccc}
\hline
Obs. ID		& Energy band  & Average CR		& $\chi^{2}_\nu$ (d.o.f)	& $P_s$  \\ 
		&		& (counts s$^{-1}$)	&				& (\%)\\ 
\hline
00033012009  	& X-ray		& $0.09 \pm 0.01$	& 4.30 (47)			& 95.8\\
		& UV	      	& $2.18 \pm 0.05$	& 5.38 (47)			&\\
00033012013  	& X-ray		& $0.12 \pm 0.01$	& 2.84 (39)			& 97.7\\
		& UV	  	& $2.28 \pm 0.06$	& 2.63 (39)			&\\
00033012020  	& X-ray		& $0.17 \pm 0.01$	& 2.85 (92)			& 99.8\\
		& UV	  	& $1.22 \pm 0.03$ 	& 2.81 (92)			&\\
\hline
\end{tabular}
\end{center}
\end{table} 

\subsection{UV data (2013 November 6--19)}
J1023 has been monitored with the UVOT in event mode using the UVW1 filter (2600 \AA) for 13 consecutive observations, 
from 2013 November 6--19 (Obs. ID 00033012003--00033012016). We processed these data with the \textsc{coordinator} and 
\textsc{uvotscreen} tasks and we extracted all the light-curve data with \textsc{xselect} (v. 2.4), adopting as extraction region a 
circle centred on the source with a radius of 10 pixels (1 UVOT pixel = 0.502 arcsec). Folding the UV light-curve to the 
4.75-h orbital period, we derive a $3\sigma$ upper limit of $\sim 0.11$ counts to any sinusoidal modulation.

\subsection{X-ray/UV correlations}
To search for possible correlations between the count rates in the X-ray and UV bands on time-scales of days, we extracted 
the XRT events in the November 6--19 period, using as extraction region a circle centred on the source with a radius of 15 
pixels (1 XRT pixel = 2.36 arcsec). The 0.3--10 keV and UVW1 light-curves of J1023 during this period are shown in Fig. \ref{corr}.
We found a strong correlation between the X-ray and UV light-curves, with a significance probability of 99.6 per cent according to 
Spearman's test. We also investigated if such correlation exists on shorter time-scales by analysing each of the three observations 
with the longest exposures (Obs. ID 00033012009, 00033012013, 00033012020). We found that both the X-ray and the UV emissions 
are variable during each observation and that the correlation is always significant (see Table \ref{varcorr}).

\subsection{X-ray/UV power spectra}
We searched for possible X-ray and/or UV periodicities by inspection of the {\it Swift} XRT and UVOT observations performed 
between 2013 November 6 and 19. We note that due to the {\it Swift}/XRT sensitivity, the time resolution does not allow 
the detection of pulsations at the pulsar spin period. First, we applied barycentric corrections to each event file with the \textsc{barycorr} task, 
using the DE-200 Solar system ephemeris. We then summed all the event files and built the power spectra both in the 0.3--10
keV band and in the UVW1 filter with \textsc{powspec}. No prominent features can be observed, implying that the X-ray and UV emissions
are not modulated at the 4.75-h orbital period (see also Tendulkar et al. 2014).

\section{Optical and infrared photometry}
\label{optphot}
J1023 was monitored in the optical and NIR bands with the REM telescope (Zerbi et al. 2001; Covino et al. 2004) at the 
La Silla Observatory on 2013 November 17 and 30, and in the optical with the 2-m Liverpool Telescope (LT) on 2014 
February 2.

\begin{figure}
\begin{center}
\includegraphics[width=5.4cm,angle=-90]{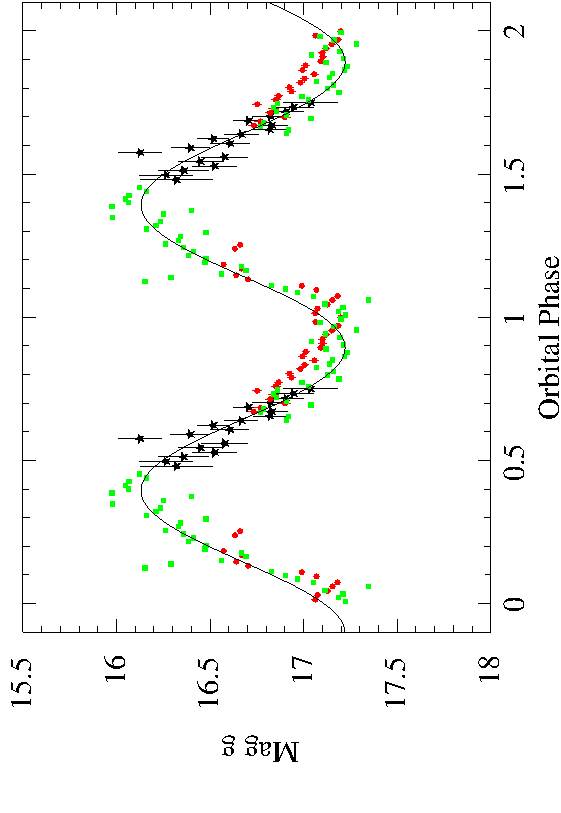}
\includegraphics[width=5.4cm,angle=-90]{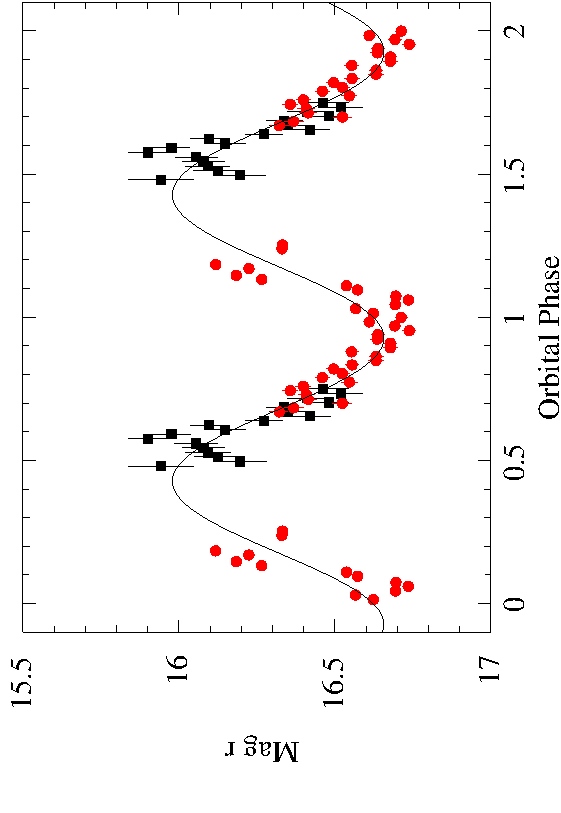}
\includegraphics[width=5.4cm,angle=-90]{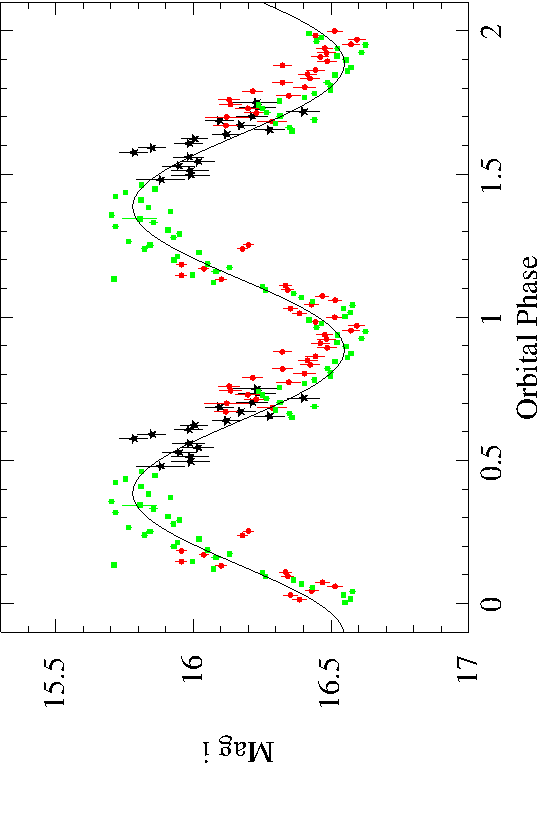}
\caption{Optical light-curves of J1023. Black stars and red dots refer to the REM 2013 November 17 and 30 data sets 
respectively; green squares refer to the LT data set. We calculated the orbital phases based on the ephemeris of 
Archibald et al. (2009). Phase 0 corresponds to the inferior conjunction of the companion. The $g$ and $i$ LT curves 
were rescaled to the mean REM 2013 Nov 30 magnitude, by adding 0.22 mag to the $g$ LT light-curve data
and 0.17 mag to the $i$ LT light-curve data. Data were fitted with a constant plus sinusoid model, with the sinusoidal 
period fixed to the orbital period of the system (4.75 h). Two cycles of the system are drawn for clarity in each plot. 
Top: $g$ light-curve; middle: $r$ light-curve; bottom: $i$ light-curve.}
\label{lc}
\end{center}
\end{figure}

\begin{figure}
\begin{center}
\includegraphics[width=8.8cm]{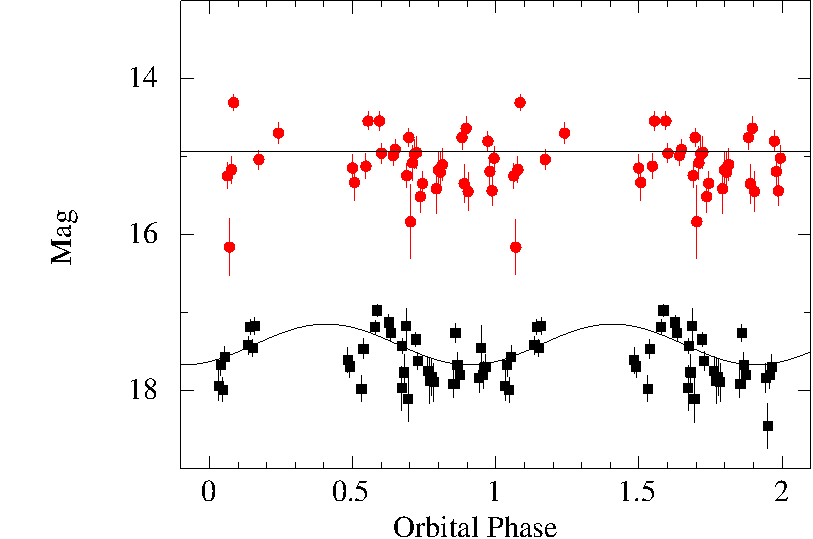}
\caption{$H$ (red dots) and $J$ (black squares) light-curves for the system J1023. The $J$-band light-curve has been 
   rescaled by 2 mag to eliminate any overlap of the two light-curves. Two cycles are drawn for clarity.}
\label{lc_nir}
\end{center}
\end{figure}

\begin{figure}
\begin{center}
\includegraphics[width=5.6cm,angle=-90]{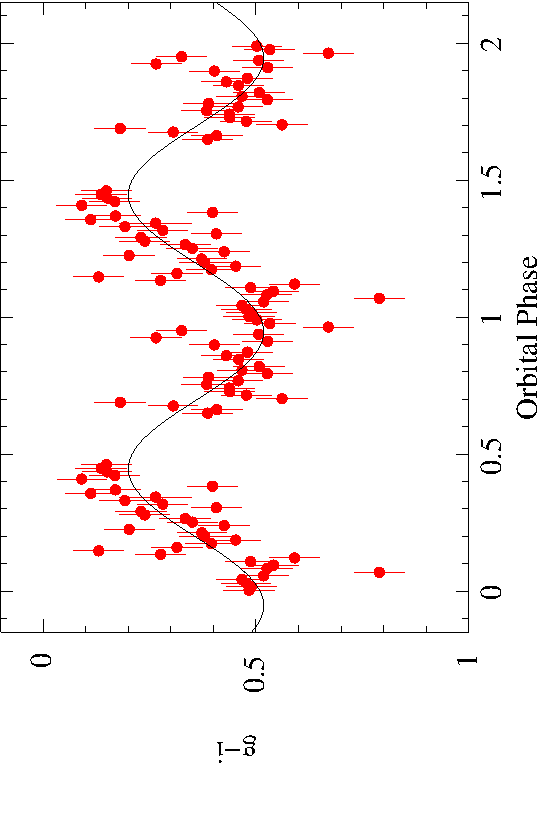}
\caption{$g$-$i$ folded colour light-curve obtained from the LT data. Fitting the data with a constant plus sinusoid
model yields an average colour $g$-$i=0.36\pm0.01$, a semi-amplitude of $0.16\pm0.02$ and an orbital phase for the maximum of 
$0.39\pm0.001$ ($\chi^{2}=178.9$ for 57 d.o.f.).}
\label{g_i}
\end{center}
\end{figure}

\begin{table*}
\caption{Results of the optical and NIR photometry of J1023. Magnitudes are not corrected for reddening, whose parameters are
reported in the last column and are derived from Cox (2000). Errors are quoted at a 90 per cent confidence level.}             
\label{phot}      
\centering                          
\begin{tabular}{cccccc}        
\hline               
Filter & Telescope/Instrument 	& Semi-amplitude 	& Mean magnitude	& Maximum 		& $A_{\lambda}$\\     
       &			& (mag)	         	& (mag)        		& (phase)   		&	\\ 
\hline                        
$g$ 	& REM/ROSS2	& $0.43 \pm 0.01$ 	& $16.70 \pm 0.01$ 	& $0.440 \pm 0.002$ 	& $0.21 \pm 0.05 $ \\
        & LT/IO:O 	& $0.55 \pm 0.01$	& $16.447 \pm 0.002$ 	& $0.393 \pm 0.001$ 	& \\					
$r$ 	& REM/ROSS2	& $0.34 \pm 0.01$ 	& $16.32 \pm 0.01$ 	& $0.427 \pm 0.003$ 	& $0.14 \pm 0.04$ \\
$i$ 	& REM/ROSS2	& $0.30 \pm 0.01$ 	& $16.16 \pm 0.01$ 	& $0.419 \pm 0.001$ 	& $0.11 \pm 0.03$ \\
	& LT/IO:O		& $0.38 \pm 0.01$	& $15.994 \pm 0.002$	& $0.386 \pm 0.001$	& \\
$J$ 	& REM/REMIR	& $0.26 \pm 0.05$ 	& $15.41 \pm 0.03$ 	& $0.41 \pm 0.02$ 	& $0.05 \pm 0.01$ \\
$H$ 	& REM/REMIR	& $-$ 		 	& $14.94 \pm 0.03$ 	& $-$ 			& $0.03 \pm 0.01$ \\
\hline                                   
\end{tabular}
\end{table*}

\subsection{REM observations}
The system was observed simultaneously in the $g$, $r$, $i$ and $z$ optical SDSS filters (4000--9500 \AA) using the ROSS2 
instrument. Two sets of 18 and 36 images were acquired during the two nights, respectively (150-s integration time each). The 
images were flat-field and bias corrected using standard procedures, and all the magnitude values for the objects in the 
field were obtained using aperture photometry techniques (\textsc{daophot}; Stetson 2000). The flux calibration was 
performed using five SDSS\footnote{http://sdss3.org/dr10/} stars present in the field. About 85 per cent of the orbital 
period was covered in each optical filter.

The system was observed also in the NIR band using the REMIR infrared camera. A set of 12 images were acquired with the 
$J$, $H$, $K$ filters (1-2.3 $ \mu \rm m $) during the first night; 24, 22 and 20 images were obtained in the $J$, $H$, $K$ 
filters during the second night (75-s integration time each). Magnitudes were extracted using the same procedure employed in the 
optical analysis, and were then calibrated against eight field stars whose magnitudes are tabulated in the 2MASS 
catalogue\footnote{http://www.ipac.caltech.edu/2mass/}.

\subsection{LT observations}
J1023 was observed in the $g$ (4770 \AA) and $i$ (7625 \AA) optical SDSS filters also with the IO:O instrument of the 
LT. The seeing remained almost constant below 1 arcsec for the whole night. A set of 63 images (90-s integration time 
each) were obtained in the two filters, covering about the 80 per cent of the orbital period in each band. Image reduction 
and magnitude extractions were carried out following the same prescriptions as for the REM data. The flux calibration
was performed using nine SDSS field stars.

\subsection{Results}
The optical counterpart star of J1023 is well detected in the single observations by REM in the $g$, $r$ and $i$ 
filters, but not in the $z$ band (a value of $16.2\pm 0.2$ mag is obtained for the magnitude in this band when all images are
summed together, for a total integration time of 2700 s). It is also clearly detected by the LT in the $g$ and $i$ 
filters. The light-curves present a sinusoidal modulation at the 4.75-h orbital period (see Fig. \ref{lc}). To study the 
modulation, we fitted the data with a simple sinusoidal function, although it provides only an approximation to the data. 
The results of the fit are reported in Table \ref{phot}. The source shows a single minimum around phase 0 (i.e. at the 
inferior conjunction of the companion star) and a maximum around phase 0.5 (superior conjunction), which is suggestive of a 
companion star strongly irradiated by the compact object and it is the typical variability expected (and observed) for an 
accreting millisecond pulsar (Homer et al. 2001; Campana et al. 2004; D'Avanzo et al. 2007; Deloye et al. 2008; Wang et al. 
2009b, 2013; Baglio et al. 2013). No significant flickering or flaring activity can be observed in our optical light-curves.

The NIR phase-resolved light-curves are shown in Fig. \ref{lc_nir}. The source is not detected in the single observations in 
the $K$ band (a value of $15.2 \pm 0.2$ mag is obtained for the magnitude in this band when all images are summed together, 
for a total integration time of 1350 s). The fit of the $H$-band light-curve with a constant gives $\chi^{2}_\nu \sim4$ for 
65 d.o.f., thus indicating that some variability is indeed observed for J1023 in this band. However, an $F$-test proves that 
a sinusoidal fit does not improve the significance of the fit. Therefore, we conclude that we are probably observing some kind 
of random variability around an average magnitude of $14.94 \pm 0.03$ mag (see Table \ref{phot}). In the $J$ band, the light 
curve is better described by a sinusoidal model than by a constant alone (the $F$-test gives a significance of $2.8\times 10^{-5}$).
All the fit parameters are reported in Table \ref{phot}.

The $g$-$r$ colour derived from the REM optical data and the $g$-$i$ colour obtained from the LT data set 
(see Fig. \ref{g_i}) are larger at phase 0 than at phase 0.5. This means a bluer spectrum at the superior conjunction, as 
expected for an irradiated companion star. If we suppose the companion to be a main sequence star, the unabsorbed overall 
average colours $g$-$r$ = $0.31\pm0.07$ mag and $r$-$i$ = $0.13\pm0.05$ mag derived from the REM data are marginally 
consistent with a $G$-type star (Cox 2000). We note however that the average $J$-$H$ colour ($0.45\pm0.04$ mag) is indicative 
of a much colder star (a $K$-type star). This inconsistency can be explained assuming the contribution of at least another component 
to the system emission, such as an accretion disc around the compact object and/or a shock front between matter outflowing from 
the companion and the relativistic pulsar wind. If this is verified, all the calculated magnitudes refer to a multi component emission 
and cannot give any estimate of the star surface temperature. 

\section{Optical spectroscopy}
J1023 optical spectra were obtained on 2013 Dec 4 with the 2.1-m telescope at the San Pedro M\'{a}rtir Observatory 
(M\'{e}xico) equipped with the Boller \& Chivens spectrograph. We obtained four optical spectra with exposure times of 1200 
or 1800 s, each one covering the 4000-7800 \AA\ wavelength range, with a resolution of $\sim6.5$ \AA\ (350 km s$^{-1}$). 
An additional 1800-s spectrum was obtained on 2013 Dec 10 at the Loiano Astronomical Observatory (Italy) using the 1.5-m 
telescope equipped with the BFOSC spectrograph, covering the 4000--8500 \AA\ wavelength range, with a resolution of about 10 
\AA\ (480 km s$^{-1}$). The log of the optical spectroscopy observations is shown in Table~\ref{tab_log_spe}.

Data were reduced using standard procedures for bias subtraction and flat-field correction. Wavelength calibration was carried 
out using copper-argon lamps. Instrumental flexures during our observations were then accounted for using atmospheric emission 
lines in the sky spectra. 

\begin{figure}
\begin{center}
\includegraphics[width=5.5cm,angle=-90]{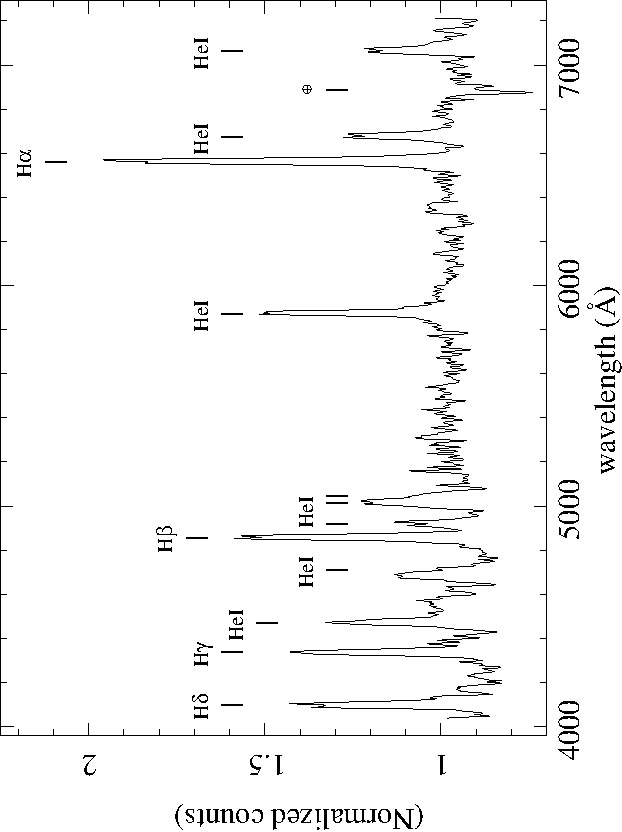}
\includegraphics[width=5.4cm,angle=-90]{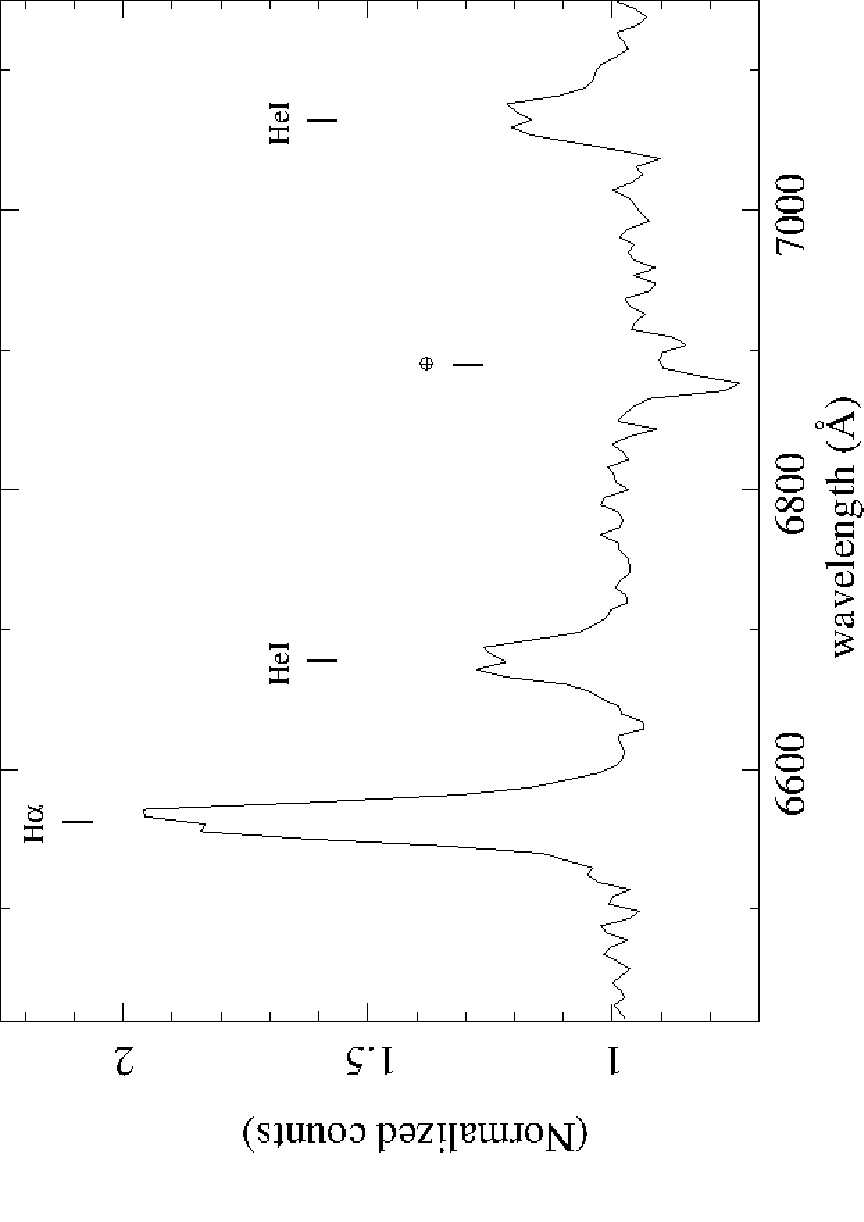}
\caption{Optical spectrum of J1023 obtained on 2013 Dec 4 with the 2.1-m telescope at the San Pedro M\'{a}rtir 
Observatory. The most prominent emission lines are marked. Double profiles are clearly visible in the lower panel.}
\label{optspec}
\end{center}
\end{figure}

The flux-calibrated spectra show a blue continuum with broad emission lines superposed. We clearly detect H$\alpha$, H$\beta$,
H$\gamma$, H$\delta$ and H$_{eI}$ ($\lambda 4472, 4713, 4921, 5016-5048, 5876, 6678, 7065$ \AA) all showing a double-horned
profile, likely related to the presence of an accretion disc (see Fig. \ref{optspec}). 

From the analysis of the co-added San Pedro M\'{a}rtir spectra, we measure an equivalent width EW $\sim -30$ \AA, a FWHM 
$\sim 1300$ km s$^{-1}$ and a peak separation of $\sim 720$ km s$^{-1}$ for the H$\alpha$ line. These values are consistent 
with the findings of Halpern et al. (2013), Takata et al. (2014) and Linares et al. (2014). We repeated the same 
analysis for the other main emission lines, excluding those with wavelength $< 4800$ \AA\ (being too close to each other 
to enable a detailed quantitative analysis). The results are reported in Table~\ref{tab_EW}. Fully consistent values are 
found from the analysis of the single spectrum obtained in Loiano.

\begin{figure}
\begin{center}
\includegraphics[width=5.5cm,angle=-90]{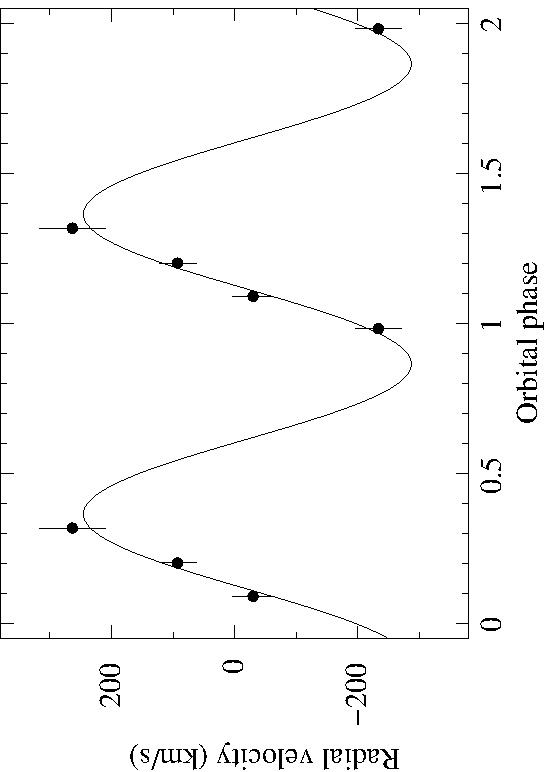}
\caption{Radial velocity curve for J1023 computed using a G5 V spectral template. Two orbital phases are shown for 
clarity. The best sine-wave fit provides a systemic velocity $\gamma = -20.6 \pm 120.8$ km s$^{-1}$ and a radial velocity 
$K_2 = 266.2 \pm 28.8$ km s$^{-1}$.}
\label{radvel}
\end{center}
\end{figure}

\begin{figure*}
\begin{center}
\includegraphics[width=8.7cm]{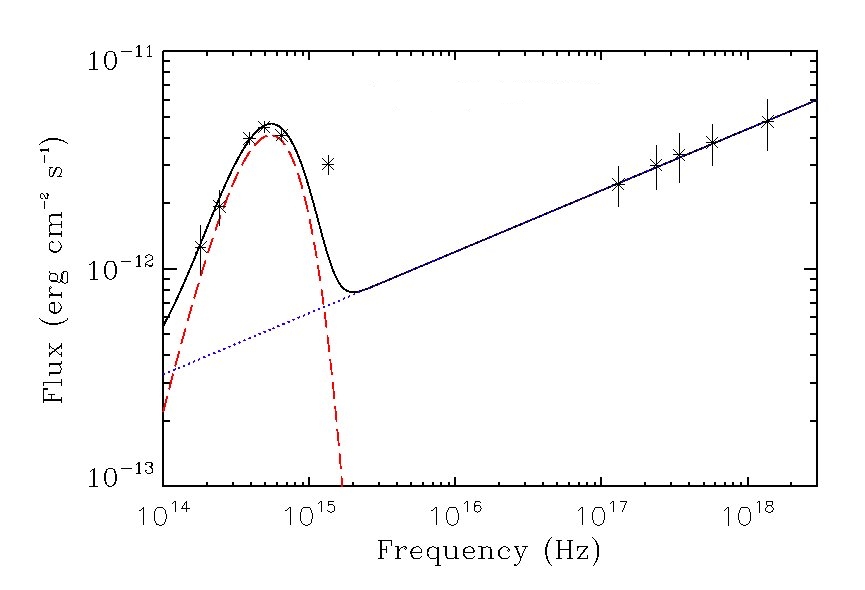}
\includegraphics[width=8.7cm]{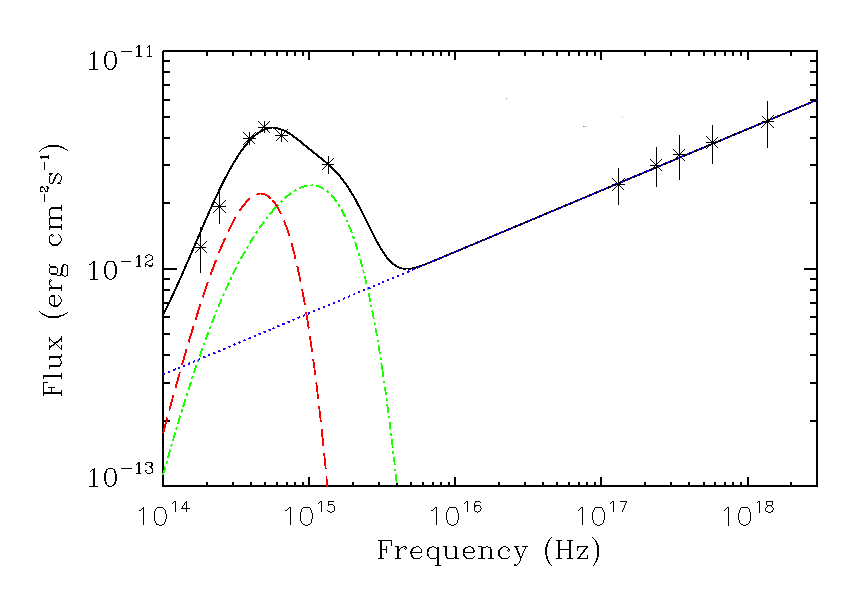}
\caption{Spectral energy distribution from NIR to X-rays obtained using data collected at the same time (within $\sim 3$ min) 
corresponding to an orbital phase of 0.9. In the left-hand panel we report the fit with a model made by the companion star (red 
dashed line) and the shock emission powered by the neutron star spin-down luminosity (blue dotted line). In the right-hand panel, 
the same data are fitted with the addition of an accretion disc contribution (green dash--dotted line).}
\label{sed}
\end{center}
\end{figure*}

\begin{table}
\scriptsize
\caption{Log of optical spectroscopy observations.}
\centering
\begin{tabular}{cccc} \hline 
Date (mid exposure)  &  Exp    &  Telescope/Observatory  &   Orbital Phase   \\
  YYYYMMDD              & (s)           &                        &                   \\ \hline
20131204.44641	  & 1800         &  2.1m/SPM               & 0.982	       \\
20131204.46790	  & 1800         &  2.1m/SPM               & 0.090	       \\
20131204.48981	  & 1800         &  2.1m/SPM               & 0.201	       \\
20131204.51290	  & 1200         &  2.1m/SPM               & 0.318	       \\
20131210.19627	  & 1800         &  1.5m/Loiano            & 0.111	       \\
\hline
\end{tabular}   
\label{tab_log_spe}
\end{table}

\begin{table}
\caption{Results of optical emission lines analysis.}
\begin{tabular}{cccc} \hline 
Line           &  EW                &  FWHM            & Peak separation      \\
               & (\AA)              & (km s$^{-1}$)    & (km s$^{-1}$)         \\ \hline
H$\alpha$      & $ -30.0 \pm 0.2$   &  $1260 \pm 110$  & $720 \pm  50$	  \\
H$_{eI}$ 5876  & $ -17.3 \pm 0.2$   &  $1300 \pm 70$   & $510 \pm 200$       	  \\
H$_{eI}$ 6678  & $  -7.9 \pm 0.2$   &  $1560 \pm 970$  & $760 \pm  60$ 	  \\
H$_{eI}$ 7065  & $  -4.9 \pm 0.2$   &  $1350 \pm 180$  & $760 \pm 170$ 	  \\
\hline
\end{tabular}   
\label{tab_EW}
\end{table}

Radial velocities were measured from the four San Pedro M\'{a}rtir optical spectra through cross-correlation with a G5 V 
template star\footnote{In order to avoid problems in comparing spectra obtained with different instruments, we did not 
include the Loiano spectrum in the radial velocity analysis. Furthermore, we note that the orbital phase at which it was 
observed is already covered by the San Pedro M\'{a}rtir spectra.}. 
Prior to cross-correlating, the spectra were barycentric corrected and rebinned to a uniform velocity scale. Cross-correlation 
was performed in the range 5130-6300 \AA, after masking the H$_{eI}$ $\lambda5876$ emission line from the accretion 
disc. A fit with a constant plus sinusoid function provides a systemic velocity $\gamma = -20.6 \pm 120.8$ km s$^{-1}$ and a 
radial velocity $K_2 = 266.2 \pm 28.8$ km s$^{-1}$ (see Fig. \ref{radvel}). Although the large uncertainties 
(mainly due to the low statistics), these values are fully consistent with the results obtained by Thorstensen \&
Armstrong (2005) during the system quiescence. 

\section{Spectral energy distribution}
J1023 was observed strictly simultaneously by {\it Swift} and REM for $\sim 3$ min on 2013 November 30, while it was at orbital 
phase 0.9 (near the neutron star superior conjunction). We extracted the corresponding source and background X-ray spectra adopting 
as extraction region a circle centred on the source with a radius of 15 pixels in the former case and a circle positioned in a location free from known 
X-ray sources with a radius of 30 pixels in the latter. We created the ancillary response file for the extracted spectrum with 
\textsc{xrtmkarf}, in order to correct the count rate for the presence of bad pixels, vignetting and hot columns. Finally, we assigned 
the latest version of the redistribution files. Fluxes in the optical and infrared bands were estimated through the aperture 
photometry technique. The unabsorbed SED from the NIR to the X-ray bands is shown in Fig. \ref{sed}.

We attempted to account for the NIR, optical, UV and X-ray SED with the simple model of an 
irradiated star plus the contribution of a shock front. For the star component we assumed an irradiated blackbody model 
(for the details of the modelling see Eqs. [8]-[9] of Chakrabarty 1998) which depends on the irradiating luminosity ($L_{irr}$), the source 
distance ($D$), the radius of the companion star ($R_c$), the albedo of the star (${\eta}_*$) and the binary separation ($a$). 

We fitted the data by using $L_{irr}$ as a free parameter and fixing $D=1368$ pc (Deller et al. 2012), 
$R_c = 0.43 \, R_{\odot}$ (Archibald et al. 2009) and ${\eta}_* = 0.1$. The binary separation 
$a~=~[G(M_X~+~M_c)(P_{\rm orb})^2/(4{\rm \pi})^2]^{1/3}$ was obtained by fixing $M_X = 1.7 \, M_{\odot}$ and 
$M_c = 0.24 \, M_{\odot}$ (Deller et al. 2012). We modelled the shock front component with a simple power law, by leaving free 
to vary the normalization and the power law index. The best fit is obtained for $L_{irr} \sim 5 \times 10^{34}$ erg s$^{-1}$, 
consistent with the estimated dipole spin-down luminosity of the ms radio pulsar 
($L_{\rm sd} = [4.43 \pm 0.04] \times 10^{34}\ergs$; Archibald et al. 2013). However, such model provides a rather poor fit to 
the data, particularly in the UV region (${\chi}^2 = 46.6$ for 10 d.o.f.; see Fig. \ref{sed}, left-hand panel).

To check for a more realistic solution, we tried to fit our data by adding to the model the contribution of an irradiated 
accretion disc by using $L_{irr}$ and the internal disc radius ($R_{in}$) as free parameters (Eqs. [10]-[15] of Chakrabarty 
1998). We fixed the X-ray albedo of the disc to 0.95 (Chakrabarty 1998) and the mass transfer rate to 10$^{-11} \, M_{\odot}$ 
yr$^{-1}$, as predicted by Verbunt (1993) for a short-period X-ray binary with a main-sequence companion star and where the 
mass transfer is kept going by loss of angular momentum from the system. For each possible radius $R_{in}$, we assumed an 
outer disc radius of $0.3a$ (where $a$ is the binary separation). The fit improves ($\Delta {\chi}^2 = 41.8$ with respect to 
the star plus shock model; see Fig. \ref{sed}, right-hand panel) and we obtain an acceptable solution for 
$L_{irr} \sim 1 \times 10^{34}$ erg s$^{-1}$ and an inner radius for the emitting region $R_{in} \sim 2 \times 10^{9}$ cm.
More sophisticated models are beyond the scope of the paper.

\section{Enshrouding of a radio pulsar}
Millisecond radio pulsars in binary systems can interact with the companion star if the orbital separation is small enough.
In recent years an increasing number of the so-called spider radio pulsars have been discovered (see Roberts 2011 for a 
review), following the first discoveries of the redback PSR J1740-5340 in the globular cluster NGC 6397 (with a non-degenerate 
companion, D'Amico et al. 2001) and the black widow PSR B1957+20 (with a brown dwarf companion, Reynolds et al. 2007).
The interaction between the relativistic pulsar wind and matter outflowing from the companion can give rise to different geometries 
depending on the relative strengths of the two components. In the case of a strong relativistic wind (high pressure), matter outflowing from the 
companion is pushed in a narrow cometary tail around it, as in the case of PSR B1957+20. In the case of a weaker relativistic 
pulsar wind (or strong matter pressure) more extended patterns can be obtained, giving rise to extensive eclipses of the radio 
signal during the orbital period. A further case has been envisaged by Tavani (1991), where a large amount of matter is 
outflowing from the companion, completely engulfing the radio pulsar. In this regime the radio pulsar is still active, but 
its signal is completely undetectable in the radio band due to the high free--free absorption.

Arons \& Tavani (1993) developed the theory of high-energy emission by the relativistic shock produced by the pulsar wind
in the nebula surrounding the binary and by the shock constraining the mass outflow from the companion star. 
Particularly compelling to J1023 is the case of PSR B1259--63. This is a millisecond pulsar orbiting a high mass Be 
companion in a very eccentric orbit. Far from the companion, the X-ray emission is dominated by shock-powered high-energy 
emission produced by the interaction between the relativistic wind from the pulsar and matter outflowing from the companion. 
The spectrum is described by a hard power law with $\Gamma\sim 1.5$, extending up to 200 keV (and more) as detected with 
COMPTON/OSSE (Grove et al. 1995). During its orbital evolution, the pulsar disappears close to periastron. During the 
passage an increase in luminosity and a softening of the power law $\Gamma\sim1.9-2$ are observed (Tavani \& Arons 1997). 
A change to a propeller regime has been excluded (Campana et al. 1995).

Three characteristic radii define the fate of matter falling on to a magnetized, fast spinning neutron star: the 
magnetospheric radius,  $r_{\rm m}$ (where the incoming matter pressure balances the magnetic dipole pressure), the 
corotation radius, $r_{\rm cor}$ (where matter in Keplerian orbit corotates with the neutron star), and the light cylinder 
radius, $r_{\rm lc}$ (where field lines attached to the neutron star rotate at the speed of light). The last two radii 
depend only on the neutron star spin and in the case of J1023 their values are $r_{\rm cor}=37$ km (for a neutron star mass 
of $1.7\msole$) and $r_{\rm lc}=81$ km. The magnetospheric radius depends on the neutron star magnetic field and on the mass 
inflow rate. It can be expressed as 
$$
r_{\rm m}=57 \mdot_{15}^{-2/7}\, M_{1.7}^{-1/7} \, R_6^{12/7} \, B_8^{4/7} \ {\rm km},
$$
where 
$\mdot_{15}$ is the mass accretion rate on to the magnetosphere in units of $10^{15}$ g s$^{-1}$, $M_{1.7}$ and $R_6$ are 
the neutron star mass and radius in units of $1.7\msole$ and 10 km, respectively, and $B_8$ is the magnetic field normalized 
to $10^8$ G (Perna, Bozzo \& Stella 2006). Due to the inclination of the magnetic moment with respect to 
the rotation axis, the magnetospheric radius can be larger up to a factor 1.5 (Perna et al. 2006).

The magnetic field of J1023 has been estimated to be $B = 9.7\times 10^7$ G (Archibald et al. 2013). Depending on the mass 
inflow rate, matter can reach the neutron star surface ($r_{\rm m}\lsim r_{\rm cor}$, accretion powered), be halted or 
strongly reduced at the neutron star magnetosphere by the propeller mechanism ($r_{\rm cor}\lsim r_{\rm m} \lsim r_{\rm lc}$,
still accretion powered) or be ejected by the pulsar pressure ($r_{\rm m}\gsim r_{\rm lc}$, spin-down powered). These regimes 
occur for lower and lower mass accretion rates. When a radio pulsar reactivates it is more difficult to quench it. This is simply due 
to the fact that the mass inflow pressure has overcome the radiation pulsar pressure throughout the binary separation 
(Campana et al. 1998; Burderi et al. 2001). This is why it is so difficult to quench a fast spinning millisecond radio pulsar when it 
reactivates. It is not easy to account for the minimal luminosity needed to quench J1023. Following Burderi et al. (2001), we can 
estimate a quenching luminosity of $\sim 5\times 10^{35}\ergs$. This is much larger than the observed X-ray luminosities and would 
suggest that the radio pulsar is still active, but unobservable. 

We note that a propeller model has been proposed to explain the emission characteristics of XSS J12270-4859 (Papitto, Torres \& Li 2014) 
and such model may apply to J1023 in its current state as well. In this model the power law emission is interpreted as resulting from
synchrotron emission at the interface between the disc and the magnetosphere. Based on energetic grounds, this model is slightly 
disfavoured by the data, despite the appealing feature of easily explaining the presence of a (truncated) accretion disc. The minimum 
luminosity expected in the propeller regime is obtained when the magnetospheric radius is close to the light cylinder radius.
In the case of J1023, this luminosity is $\sim 8\times 10^{33}$ erg s$^{-1}$, whereas the maximum luminosity expected
in the propeller regime is $\sim 3\times 10^{35}$ erg s$^{-1}$. The observed mean X-ray luminosity is $\sim 3\times 10^{33}$ erg s$^{-1}$ 
and during the flare it rises to $\sim 6\times 10^{33}$ erg s$^{-1}$. By contrast the full spin-down luminosity is$\sim 4\times 10^{34}$ erg s$^{-1}$  
Based on {\it NuSTAR} data, Tendulkar et al. (2014) estimated an average value for the X-ray luminosity which is close to the minimum luminosity expected 
in the propeller regime, whereas the peak luminosity should be in the propeller luminosity interval.

\section{Conclusions}
In this work we presented the results of 51 {\it Swift} XRT and UVOT observations of the binary millisecond radio pulsar PSR J1023+0038
carried out in the 2013 October--2014 May period and the results of optical and NIR photometry and optical spectroscopy of this system.

The X-ray spectrum of J1023 is best modelled by an absorbed power law, with a column density of $(5.2\pm0.07)\times 10^{20}$ cm$^{-2}$ and 
a photon index $\sim1.6$. The contribution from any possible soft component is negligible and no iron line features can be detected. Spectral 
changes can be accounted for by changes of the power law normalization alone.  These observational features are consistent with shock-powered 
emission produced by the pulsar/outflow interaction (Tavani \& Arons 1997) and with the enshrouding of J1023 by a large amount of matter,
material which is however not enough to quench the radio pulsar. The enshrouding is testified by the mild increase in the 
column density and by the softening of the power law photon index compared to the previous quiescent period (Archibald et al. 
2010; Bogdanov et al. 2011): the former rose from {\bf $<1$} to $\sim5 \times 10^{20}$ cm$^{-2}$, the latter changed from 
$\sim 1.2-1.3$ to $\sim1.6$ (Tavani \& Arons 1997). At the end of 2014 February, X-ray flares were detected, with a 0.3--10 keV luminosity of  
$\sim5.6\times 10^{33}\ergs$ (a factor $\sim2.2$ higher than the average X-ray luminosity). The peak luminosity of J1023 during flaring is large, 
encompassing a sizeable fraction of the spin-down luminosity and thus leaving open the possibility for emission in the propeller state (Papitto et al. 2014).

The shock-powered emission scenario can be well reconciled with both the reported switch off in the radio band and the 
increase in the gamma-ray emission. According to this, the radio pulsar signal is completely shielded by scattering and/or free--free 
absorption by a large amount of outflowing matter, whereas the increase in the gamma--ray emission can be naturally explained in 
terms of an enhancement of the shock emission (Archibald et al. 2013; Patruno et al. 2014; Stappers et al. 2014; Takata et al. 2014; 
Tendulkar et al. 2014).

A strong correlation is found between the X-ray and the UV count rates in the 2013 November 6-19 period both on time-scales 
of days and of a few tens of seconds, thus suggesting that the same emission mechanism is powering part of the X-ray and UV
emission. 

Optical and infrared photometric observations show that the companion star is irradiated by the spin-down emission of the 
radio pulsar, but also suggest that other emission mechanisms must be at work. In fact, double-peaked emission lines in 
the optical spectra indicate that an accretion disc exists around the system, as shown also from multi-wavelength 
observations perfomed since 2013 June (Patruno et al. 2014; Stappers et al. 2014; Takata et al. 2014; Tendulkar et al. 2014). 
The peak separation of the horns of the emission lines is $\sim700-800$ km s$^{-1}$. This value is consistent with those reported 
by Halpern et al. (2013), Takata et al. (2014) and Linares et al. (2014), suggesting that the disc thickness has not changed significantly 
between 2013 October and December.

The SED from the NIR to the X-ray band is well represented by a model consisting of an
irradiated companion, an accretion disc and a shock emission. The SED reveals a minimal contribution of the disc to the
X-ray emission and may indicate that accretion on to the neutron star is not occurring (see also the SED reported by
Takata et al. 2014). In fact, this is what is expected in order not to completely quench the pulsar mechanism and thus the 
relativistic wind. The SED also shows that more than half of the UV emission ($\sim 60$ per cent fractional contribution to
the total flux) is emitted from the accretion disc, with the remaining part arising from the intra-binary shock ($\sim 40$ 
per cent). The modelling of the SED allowed us to estimate the inner radius of the emitting region 
($R_{in} \sim 250$ $r_{\rm{lc}}$), consistent with the value reported by Takata et al. (2014) for the inner edge of the 
accretion disc, and the spin-down luminosity of the radio pulsar ($L_{irr} \sim 1 \times 10^{34}$ erg s$^{-1}$), possibly
involving some shielding.

If this picture is correct, the disc should not extend down to the magnetosphere, but it should be halted further outside.
The disc should then be partially supported or strongly evaporated by the neutron star spin-down luminosity, as it has been 
suggested for some white dwarfs (Meyer \& Meyer-Hofmeister 1994). Future multi-wavelength monitoring campaigns may shed 
further light on this and, more in general, on the phenomenology of the system emission.

\section*{Acknowledgements}
We thank the referee for useful comments. We thank Alessandro Patruno and Alessandro Papitto for productive discussions on 
X-ray variability and on radio pulsar-dominated and accretion-dominated regimes. SC thanks Neil Gehrels for approving {\it Swift} 
observations of J1023.We thank San Pedro M\'{a}rtir staff for all their help during the observation run. We thank Silvia Galleti for 
Service Mode observations at the Loiano telescope, and Antonio De Blasi for night assistance. TMD acknowledges funding via 
an EU Marie Curie Intra-European Fellowship under contract no. 2011-301355. EJB and HOF acknowledge CONACYT GRANT 
129204. HOF is funded by a postdoctoral UNAM grant. FGRF is funded by a CONACYT grant for postgraduate studies. This work 
made use of data supplied by the UK {\it Swift} Science Data Centre at the University of Leicester and was funded in part by European 
Research Council Advanced Grant 267697 `4${\rm \pi}$ sky': Extreme Astrophysics with Revolutionary Radio Telescopes.

\label{lastpage}

\end{document}